\documentstyle[aps,pre,twocolumn,psfig,floats]{revtex}
\begin{document}
\wideabs{
\title{Signatures of granular microstructure in dense shear flows}
\draft
\author{Daniel M. Mueth$^1$, Georges F. Debregeas$^1$, Greg S. Karczmar$^2$, Peter J. Eng$^3$,\\ Sidney R. Nagel$^1$, and Heinrich M. Jaeger$^1$}

\address{$^1$The James Franck Institute and Department of Physics,
	The University of Chicago,\\
	5640 S. Ellis Avenue, Chicago, IL 60637\\
        $^2$Radiology Department, The University of Chicago \\
        $^3$Center for Advanced Radiation Sources(CARS), The
	University of Chicago}

\maketitle

\bibliographystyle{prsty}

 

%
}

{\bf Granular materials react to shear stresses differently than do
ordinary fluids.  Rather than  deforming uniformly, materials such as
dry sand or cohesionless powders develop shear  bands: narrow zones
containing large relative particle motion leaving adjacent regions
essentially rigid~\cite{1,2,3,4,5}.  Since shear bands mark areas of
flow, material failure  and energy dissipation, they play a crucial
role for many industrial, civil engineering and  geophysical
processes~\cite{6}.  They also appear in related contexts, such as in
lubricating  fluids confined to ultra-thin molecular layers~\cite{7}.
Detailed information on motion {\it within} a shear band in a
three-dimensional geometry, including the degree of particle  rotation
and inter-particle slip, is lacking.  Similarly, only little is known
about how  properties of the individual grains - their microstructure
- affect movement in densely  packed material~\cite{5}.  Combining
magnetic resonance imaging, x-ray tomography, and  high-speed video
particle tracking, we obtain the local steady-state particle velocity,
rotation and packing density for shear flow in a three-dimensional
Couette geometry.   We  find that key characteristics of the granular
microstructure determine the shape of the  velocity profile.  
}
 
In order to probe the role of microstructure inside the narrow
granular shear zone,  independent determinations of the velocity and
density profiles with spatial resolution well  below the size of
individual particles are required.  Non-invasive measurements of this
type so  far have been limited to two-dimensional (2D) geometries
where optical tracking of all particle  positions is
straightforward~\cite{2,4,8,9,10,11}.  In a 3D Couette cell, as
sketched in Fig.~1a, a  steady-state shear flow can be set up by
confining granular material between two concentric,  vertical
cylinders and turning the inner cylinder at constant velocity $v_{\rm wall}$
while keeping the outer  wall at rest.  Unlike 2D Couette
cells~\cite{9,10}, where particles are confined to a single layer
with constant volume, there is a free upper surface allowing the
packing density to adjust via  feedback between shear-induced dilation
and gravity.

\begin{figure}
\centerline{ \psfig{file=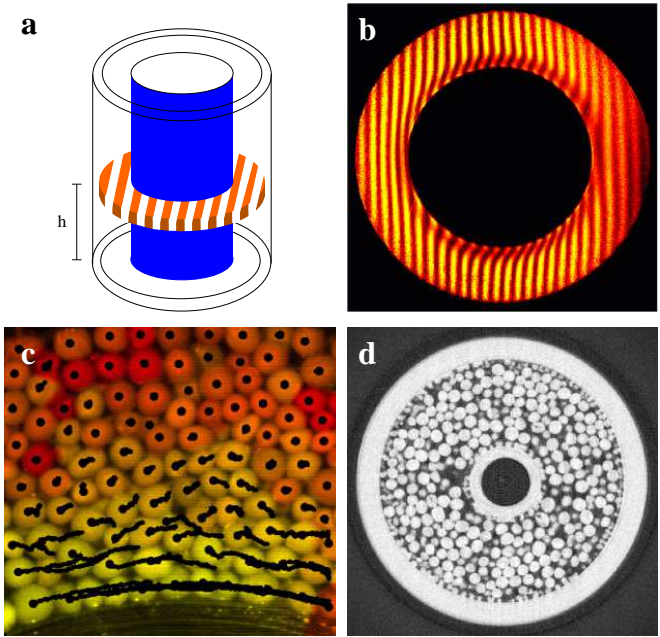,width=3.43in}}
\vspace{2ex}
\caption{ Non-invasive MRI, x-ray tomography and high-speed video
probes of granular Couette flow. (a) Sketch of the Couette-type shear
cell consisting of two concentric cylinders with diameters $51{\rm mm}$ and
$82{\rm mm}$ and filled to a level of $60{\rm mm}$.  The inner cylinder was rotated at
angular velocities from 0.6-45rpm, corresponding to $0.6{\rm mm/s}
< v_{\rm wall} < 120{\rm mm/s}$.  The flow velocity was measured using
a MRI spin-tagging technique~\protect\cite{14,15}.  Before imaging,
proton spins were encoded (spin-tagged) so as to display parallel
stripes when imaged.  Images were taken of $5{\rm mm}$ thick horizontal slices
at various heights, $h$.  (b) When the inner cylinder was rotated,
distortion of the stripe pattern revealed the displacement of the
material that occurred during the 100ms interval between spin-tagging
and imaging.  2048 spin-tag-image steps were used to assemble each
complete image.  (c) High-speed video frame, taken at 1/1000s, of
mustard seeds observed through the cell's transparent bottom.  To
indicate the movement history of individual particles, their center
positions over the preceding 200 frames are traced by black lines. The
particle coloring reflects the magnitude of each particle's average
velocity during this 0.2s interval: fast particles (yellow) near the
inner wall appear to move smoothly, while slower particles (orange and
red) display more irregular and intermittent motion.  In the long-time
average, measured by the MRI technique, the flow appears smooth
everywhere. (d) Two-dimensional, horizontal slice through the cell at
half the filling height, taken from x-ray tomography data set.
Particles, in this case mustard seeds, appear bright (as do the walls
of the cell).  These measurements were performed at the Advanced
Photon Source at Argonne National Lab in a Couette cell of slightly
smaller size but under otherwise identical shearing conditions.}
\label{fig:apparatus}
\end{figure}

The difficulty of imaging the interior has restricted studies of 3D
granular Couette  systems to either probing only the
surface~\cite{12}, to tracking colored tracers in very narrow  (few
particles wide) cells~\cite{13}, or to measuring global quantities,
such as the total applied  torque and its fluctuations~\cite{12}.
Here we use magnetic resonance imaging (MRI) to obtain  flow
velocities from the interior of a 3D system~\cite{14,15}.  We have
used oil-rich seeds as a  source of free protons that can be traced
using MRI~\cite{14}.  Two kinds of seeds were used to  explore the
role of microstructure: mustard seeds (spherical with mean diameter
${\rm d}=1.8{\rm mm}$) and  poppy seeds (kidney-shaped with mean
diameter ${\rm d}=0.8{\rm mm}$).  The wall friction was controlled by
gluing a layer of seeds on both cylinders.  Using a spin-tagging
technique, horizontal slices were imaged, as sketched in Fig.~1a.  In
the resulting MRI image (Fig.~1b), the shear band shows up as  the
narrow region of deformed stripes near the inner, moving, cylinder.
Imaging slices at  different heights, $h$, we measured these
deformations from which the azimuthal velocity profiles  were
calculated throughout the cell.  Similarly, three-dimensional x-ray
tomography allowed us  to calculate packing fraction profiles at
various heights (Fig.~1d).  At the transparent cell bottom  additional
velocity and packing fraction information was gathered by direct,
high-speed-video  particle tracking (Fig.~1c).

Prior to each set of measurements, the cell was run until steady-state
was reached (as  determined by the fact that the packing density
profiles became stationary).  Over a total  acquisition time of
typically 17min per MRI slice, high-resolution images revealed the
long-time  average local displacements during a time interval, $\delta
t = 100$~ms (Fig.~1b).  The azimuthal velocity  at a given distance
$r$ from the inner wall was obtained by exploiting the cylindrical
symmetry of the Couette geometry:  we calculated the MRI intensities
along a circle of radius $r$, $I_{\rm turning}(\Theta)$ and  $I_{\rm
stopped}(\Theta)$, for the cell turning and at rest, respectively.  The
position of the central peak in the  cross-correlation between $I_{\rm
turning}(\Theta)$ and $I_{\rm stopped}(\Theta)$ corresponds to the
average azimuthal distance  traveled by the material during $\delta
t$.  This technique yielded the angle- and time-averaged radial
profile of the steady-state azimuthal mass flow velocity, $v(r)$,
resolving $v(r)$ to within 0.1mm/s  and $r$ to within 0.1mm.  Note
that this high-resolution mass flow velocity is the average velocity
of all material at radius $r$ and thus not only contains information
about particle translation and spin, but also depends on packing
fraction.  This differs from the average velocity of particle centers
that typically is obtained by video particle tracking techniques.

We found $v(r)$ highly reproducible from run to run even though on
shorter time scales  there are rapid velocity fluctuations from point
to point within the material that, at the  boundaries, can be observed
with high-speed video (Fig.~1c)~\cite{16}.  Within the resolution of
our measurements, $v(r)$ for both types of seeds did not vary with
height, including the regions  near the top and bottom surfaces (Note
that this indicates that non-azimuthal, secondary flow  was small and
did not affect $v(r)$, in accordance with tracer-bead
studies~\cite{13}).   Aside from  an overall scale factor, we also did
not detect any shear-rate dependence to the velocity profile  over the
entire range $5{\rm mm/s}  < v_{\rm wall} < 120{\rm mm/s}$ explored
with MRI.  (With video imaging we  verified rate independence down to
velocities as small as 0.6~mm/s)~\cite{16} .

For (nearly) monodisperse smooth, spherical particles, the decay of
$v(r)$ away from the  shearing wall is dominated by abrupt drops at
integer multiples of $r/{\rm d}$ as shown in Fig.~2a for  mustard
seeds.  In the normalized velocity gradient $\gamma(r) \equiv {{\rm d}\over
v}{\partial v \over \partial r}$ (Fig.~2b) these drops
appear as deep narrow valleys.  They are correlated with pronounced
oscillations in the packing density,  $\rho(r)$, which signal the
presence of well-defined, single-grain-wide layers near the moving
wall (Fig.~2c).  (We note that the average density approaches the
random close packing value at larger  $r$ in a manner that is
consistent with an exponential form~\cite{10}.)  The shear-induced
layering  is reminiscent of that seen~\cite{17} in the collisional
regime of dilute flows, but occurs here in  the high-density limit of
rate-independent, frictional flow.  Since the velocity drops are
highly  localized, we associate them with slipping at the interface
between adjacent layers.  Either  directly from $v(r)$ or by comparing
the integrated areas of each valley in $\gamma (r)$ we find that
across  each slip zone the velocity decreases by approximately the
same factor, ${\rm b} = 0.36 \pm 0.13$.

In addition to slip between layers, MRI resolves a non-zero velocity
gradient $\gamma(r)$ within  each layer.  This can be caused by
either particle rotation or disorder in the layering along the  radial
direction.  (Along the azimuthal and axial directions, particles
within layers certainly show  packing disorder as can be seen from the
maxima in $\rho(r)$ in Figs.~2c and 3c whose values lie significantly
below those for a crystalline configuration).  For perfectly arranged
layers, $\gamma(r)$ at  the layer centers would be determined solely
by particle rotation (spin) within each layer.  The presence of
particles which do not lie perfectly within the layers (revealed by
the non-zero values  of $\rho(r)$ between layers in Fig.~2c and seen
directly from the tracks in Fig.~1c)  reduces the gradient  in the
slip region and increases it at the layer centers.  In the limit of
complete disorder, we  would expect the staircase shape of $v(r)$ to
vanish completely.  Using information about the density, $\rho_c(r)$,
and velocity, $v_c(r)$, of the particle centers, as determined by the
high-speed video  and x-ray experiments, we find that approximately
90\% of $\gamma (r)$ at the layer centers is due to the  radial
disorder.  
In order for the MRI and high-speed video experiments to
give the same $\gamma (r)$,  particle spin must be considered,
revealing the spin profile seen in the inset of Fig.~2a.  We note that
the spin is small, increasing slowly with distance from the shearing
wall to a value at $r/{\rm d} =  4.5$ of less than one full particle
rotation per $13{\rm d}$ translation.

\begin{figure}[h]
\centerline{ \psfig{file=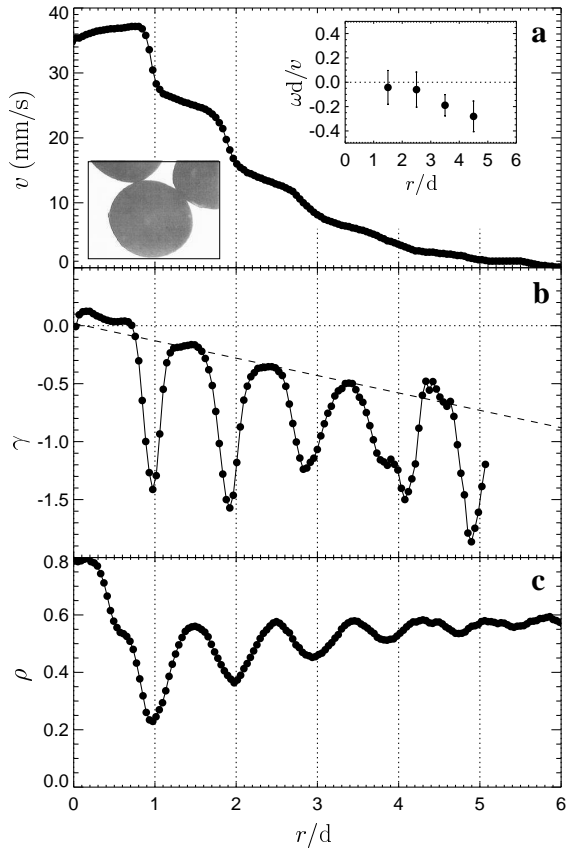,width=3.0in}}
\vspace{2ex}
\caption{  Radial velocity, spin and packing density profiles for
spherical mustard seeds. (a) Steady-state, angle-averaged azimuthal
velocity $v(r)$ across the shear band at $h = 30 {\rm mm}$, halfway
below the filling level.  The layer of seeds glued to the inner
cylinder wall extends to the dotted vertical line at $r/{\rm d}= 1$.
The lower inset shows a photograph of the seeds.  The upper inset
shows the normalized particle spin rate, $\omega{\rm d}/v$, as a
function of distance from the moving wall (where $\omega$ is the
angular velocity of a particle about its center).  (b) Normalized
velocity gradient $\gamma \equiv ({\rm d}/v) \partial v/{\partial r}$ for
the data in the main panel of (a).  (c) Angle-averaged, steady-state
radial packing density profile $\rho(r)$ computed from x-ray
tomography data as in Fig.~1d, measuring the volume fraction occupied
by seed material.   For $0 < r/{\rm d} < 1$ the seeds glued to the
wall contribute to $\rho$.}
\label{fig:mustard}
\end{figure}

The message of Fig.~2 is that the overall shape of $v(r)$ across the
shear band can be  understood as arising from two main contributions
to the velocity gradient $\gamma (r)$:  a slip  contribution in the
presence of layering, and a second contribution associated with radial
disorder.  These two pieces are distinguished by the characteristic
$r$-dependence they produce in  $v(r)$ on length scales larger than a
single grain.  A constant slipping fraction,$$\int\limits_{\rm interface}
\gamma_{\rm slip}(r)dr=-{\rm b},$$  across each interface, by itself,
leads to a velocity profile with an exponential decay, $v(r) =  v_0
\exp(-{\rm b}r/{\rm d})$.  The constant inter-layer slip appears on a
background due to radial disorder,  $\gamma_{\rm disorder}(r) = -
2{\rm c}(r - r_0)/{\rm d}$, that starts at $r_0$ within the glued-on layer
and increases linearly in strength with slope $2{\rm c}$ as indicated
by the dotted line (Fig.~2b).  Such linear $r$-dependence in
$\gamma(r)$ results in a Gaussian profile, $v(r) = v_0 exp[-{\rm
c}(r/{\rm d} - r_0/{\rm d})^2]$.  The sum of the two contributions to
$\gamma(r)$, after integration, leads to a product of an exponential
and Gaussian term for the overall,  averaged velocity profile: $$v(r) =
v_0 exp\Biggl[-{\rm b}\Bigl({r\over{\rm d}}\Bigr)-{\rm c}\Bigl({r-r_0\over{\rm d}}\Bigr)^2\Biggr]$$.  Fits
of this function to mustard seed  data yield ${\rm b}= 0.36 \pm 0.13$,
${\rm c} = 0.06 \pm 0.03$ and $r_0/{\rm d} = 0.6 \pm 0.8$ independent
of height and shear rate.

\begin{figure}[t]
\centerline{ \psfig{file=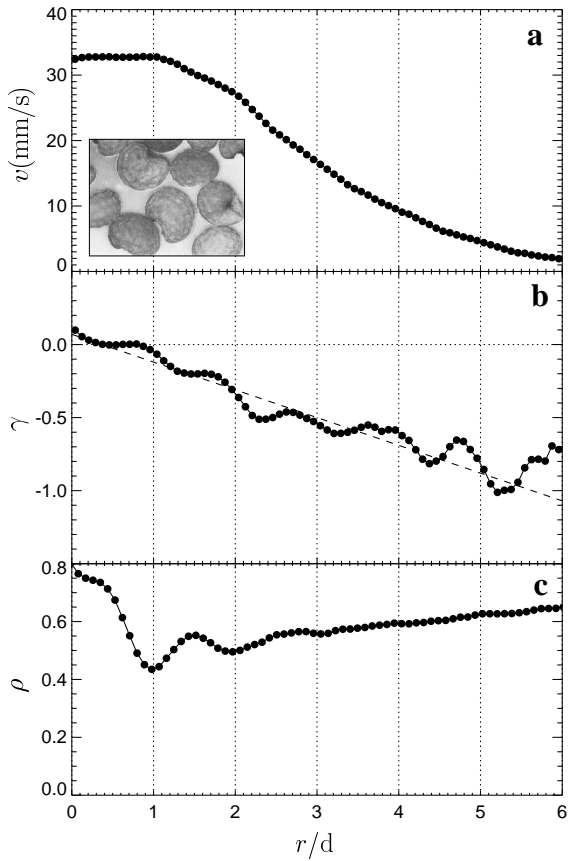,width=3.0in}}
\vspace{2ex}
\caption{ Radial velocity, spin and packing density profiles for
aspherical poppy seeds. (a) Steady-state, angle-averaged azimuthal
velocity $v(r)$ across the shear band at $h = 30{\rm mm}$.  The inset
shows a photograph of the seeds.  (b) Normalized velocity gradient for
the data in (a).  (c) Corresponding angle-averaged, radial packing
density profile $\rho(r)$ from x-ray data.  For $0 < r/{\rm d} < 1$
the seeds glued to the wall contribute to $\rho$.  }
\label{fig:poppy}
\end{figure}

The decomposition of $\gamma(r)$ allows the quantitative tracking of
the slip inside the shear  band for different particle types.  We find
a greatly reduced slip rate b when smooth spherical  particles are
replaced by roughened ones~\cite{16}.  More dramatic differences
result from  changes in the particle shape.   For kidney-shaped poppy
seeds we observe a comparatively  smooth overall velocity profile
(Fig.~3a).  Much smaller modulations in $\gamma(r)$(Fig.~3b) together
with little layering in $\rho(r)$ (Fig.~3c) indicate that interlayer
slip is greatly reduced.  The linear  trend $\gamma(r) = -2{\rm
c}(r/{\rm d} - r_0/{\rm d})$ is still present, however, and leads to
the Gaussian shape for $v(r)$ seen  in Fig.~4 over several orders of
magnitude.  These data also demonstrate explicitly that the shear rate
fixed by $v_{\rm wall}$, while setting the overall scale, leaves the shape of
the profile unchanged.  This allows for a collapse of all poppy seed
data for different shear rates and heights in the cell onto a single
Gaussian described by ${\rm c} = 0.11 \pm 0.02$ and $r_0/{\rm d} =
-0.1 \pm 0.5$.  Layering can also be suppressed  and radial disorder
be promoted in sphere packings if wide particle size distributions are
used.   We find that these, too, produce essentially Gaussian velocity
profiles~\cite{16}.

From these results, the Gaussian component of the velocity profile
emerges as a robust,  generic feature; slip, if it occurs, is seen to
augment $v(r)$ by an additional, exponential factor.  In  contrast to
the translationally invariant slip, the explicit dependence of the
Gaussian on the  distance from the shearing wall indicates
correlations.  There are several possible mechanisms  for radial
correlations~\cite{5,9,10,18,19,20}, including the formation of stress
chains~\cite{5,9,10} or particle clusters~\cite{18,20}, both of which
require high packing  densities and the absence of slip or some degree
of interlocking.  However, a detailed mechanism  that would explain
the observed Gaussian dependence presently does not exist.  A
successful  theory must not only account for the linear increase in
the magnitude of the velocity gradient with $r$, but also show how the
Gaussian shape is independent of the presence of layering.  We
speculate that both of these might be provided by a coupling of
$\gamma(r)$ to the average packing  fraction and its gradients (for
2-D Couette systems a coupling to the average density alone has  been
suggested~\cite{9,10}).  A Gaussian, then, arises naturally for a
coupling of type $\gamma(r) \propto  \rho(r_0) - \rho(r)$, considering
first order terms in $r$.  Because all stress-loading is driven by the
seed  layer glued to the inner wall, its location, $r_0$, serves as a
spatial reference point.  This argument  carries over to the absence
of a Gaussian contribution to $v(r)$ in less dense types of flow, where
interlocking may be less effective, such as rapid flows down inclines
with free upper surface for  which power law forms have been
reported~\cite{4,14,21}, or vertical gravity-driven flows  through 2D
``pipes'', which appear to follow essentially exponential
profiles~\cite{2,22}.

\begin{figure}
\centerline{ \psfig{file=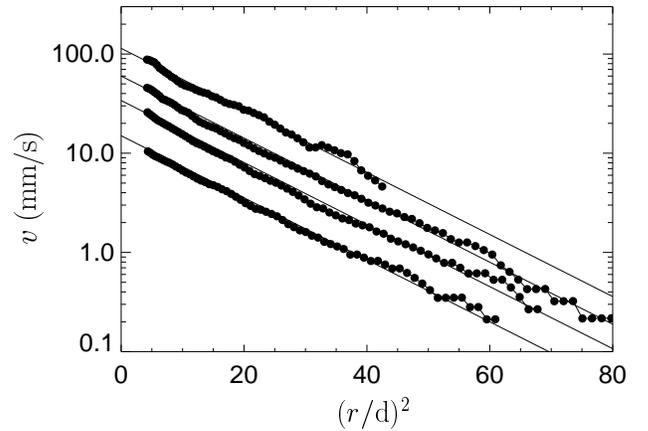,width=3.3in}}
\vspace{2ex}
\caption{ Rate-independence of the velocity profiles.  Log plot of
poppy seed velocity profiles $v(r)$ as a function of $(r/d)^2$ for
four different shearing wall speeds ($v_{\rm wall}$ = 120, 52, 33, and
14mm/s, top to bottom).  Straight lines on this plot correspond to a
Gaussian with origin at $r = 0$.  The slope, ${\rm c}$, of these lines
corresponds to half the average slope of the velocity gradient in
Fig.~3b and characterizes the width $\sigma = {\rm d}/\sqrt{{\rm c}}$
of the shear band.  }
\label{fig:rates}
\end{figure}

Our results show that in this high packing density and slow shearing
rate regime there is a  direct connection between the microstructure
and the shape of the velocity profile.  For equal-sized spherical
particles, where considerable layering is found, a strong exponential
contribution  is observed.  For aspherical seeds, which show no
pronounced layering, the profile is almost  completely described by a
Gaussian centered on the shearing wall.  Remarkably, despite the
complicated nature of the grain-grain interactions, the steady-state
behavior exhibits robust  behavior: the {\it shape} of the velocity profiles
characterized by the two length scales, $\lambda = {\rm d}/{\rm b}$
and $\sigma = {\rm d}/{\sqrt{{\rm c}}}$, is found to be height and
shear-rate independent.

\section*{Acknowledgments}
We thank Eiichi Fukushima, James Jenkins, \linebreak Christophe Josserand, Dov
Levine, Milica Medved, Vachtang Putkaradze, Mark Rivers, and Alexei
Tkachenko for helpful discussion, and Doris Stockwell from Spiceland
for the donation of mustard seeds for the experiment.  This work was
supported by an NFS research grant and by the MRSEC Program of the NSF.


\begin{references}

\bibitem{1}J. Bridgwater. On the width of failure zones. {\it Geotechnique} {\bf 30}, 533 
(1980).
\bibitem{2}R. M. Nedderman and C. Laohakul. The Thickness of the Shear Zone of Flowing 
Granular Materials. {\it Powder Technology} {\bf 25}, 91-100 (1980).
\bibitem{3}H.-B. M\"uhlhaus and I. Vardoulakis. The thickness of shear bands in granular 
materials. {\it Geotechnique} {\bf 37}, 271-283 (1987).
\bibitem{4}T. G. Drake. Structural Features in Granular Flows. {\it Journal of Geophysical 
Research} {\bf 95}, 8681-8696 (1990).
\bibitem{5}M. Oda and H. Kazama. Microstructure of shear bands and its relation to the 
mechanism of dilatancy and failure of dense granular material. {\it G\'eotechnique} {\bf 48}, 
465-481 (1998).
\bibitem{6}D. R. Scott. Seismicity and stress rotation in a granular model of the brittle crust. {\it 
Nature} {\bf 381}, 592-595 (1996).
\bibitem{7}S. Grannick. Soft Matter in a Tight Spot. {\it Physics Today} {\bf 52}, 26-31 (1999).
\bibitem{8}U. T\"uz\"un and R. M. Nedderman. An Investigation of the Flow Boundary During 
Steady-State Discharge from a Funnel-Flow Bunker. {\it Powder Technology} {\bf 31}, 27-43 (1982).
\bibitem{9}D. Howell, R. P. Behringer and C. Veje. Stress Fluctuations in a 2D Granular 
Couette Experiment: A Continuous Transition. {\it Physical Review Letters} {\bf 82}, 5241-5244 (1999).
\bibitem{10}C. T. Veje, D. W. Howell and R. P. Behringer. Kinematics of a two-dimensional 
granular Couette experiment at the transition to shearing. {\it Physical Review E} {\bf 59}, 
739-745 (1999).
\bibitem{11}J. Rajchenbach. Granular flows. {\it Advances in Physics} {\bf 49}, 229-256 
(2000).
\bibitem{12}W. Losert and J. P. Gollub. Flow properties of granular matter fluidized by gas or 
shear. {\it Bull. Am. Phys. Soc.} {\bf 44}, 47 (1999).
\bibitem{13}R. Khosropour, J. Zirinsky, H. K. Pak and R. P. Behringer. Convection and size 
segregation in a Couette flow of granular material. {\it Physical Review E} {\bf 56}, 4467-
4473 (1997).
\bibitem{14}E. Fukushima. Nuclear Magnetic Resonance as a Tool to Study Flow. {\it Annu. 
Rev. Fluid Mech.} {\bf 31}, 95-123 (1999).
\bibitem{15}E. E. Ehrichs, H. M. Jaeger, G. S. Karczmar, J. B. Knight, V. Y. Kuperman and S. 
R. Nagel. Granular Convection Observed by Magnetic Resonance Imaging. {\it Science} {\bf 
267}, 1632-1634 (1995).
\bibitem{16}D. M. Mueth, G. F. Debregeas, P. J. Eng, G. S. Karzcmar, S. R. Nagel and H. M. 
Jaeger. to be published. 
\bibitem{17}M. A. Hopkins, J. T. Jenkins and M. Y. Louge. On the Structure of Three-
Dimensional Shear Flows. {\it Mechanics of Materials} {\bf 16}, 179-188 (1993).
\bibitem{18}G. Debregeas and C. Josserand. A self-similar model for shear flows in dense granular material. {\it cond-mat 9901336} (preprint).
\bibitem{19}A. Tkachenko and V. Putkaradze. Mesoscopic physics of granular flows. {\it cond-mat/9912187} (preprint).
\bibitem{20}C. Josserand. A 2D asymmetric exclusion model for granular flows. {\it Europhysics Letters} {\bf 48}, 36-42 (1999).
\bibitem{21}D. M. Hanes and D. L. Inman. Observations of rapidly flowing granular-fluid 
materials. {\it Journal of Fluid Mechanics} {\bf 150}, 357-380 (1985).
\bibitem{22}O. Pouliquen and R. Gutfraind. Stress Fluctuations and Shear Zones in Quasi-Static 
Granular Chute Flows. {\it Physical Review E} {\bf 53}, 557-561 (1996).

\end{references}
\end{document}